\documentclass[preprint]{elsarticle}

\usepackage{amsmath,euscript,amssymb,epsfig,amsfonts,latexsym}
\usepackage{graphicx,color}

\newcommand{\be}[1]{\begin{equation}\label{#1}}
\newcommand{\ee}{\end{equation}}
\newcommand{\ba}[1]{\begin{eqnarray}\label{#1}}
\newcommand{\ea}{\end{eqnarray}}
\newcommand{\rf}[1]{(\ref{#1})}

\begin{document}

\begin{frontmatter}

\title{A quantitative analysis of the effect of box size in N-body simulations of the matter power spectrum}

\author[a]{Maxim Eingorn}

\author[b,c] {Ezgi Yilmaz}
\author[d]{A.~Emrah Y\"{u}kselci}
\author[b,c,e]{Alexander Zhuk\corref{mycorrespondingauthor}}
\cortext[mycorrespondingauthor]{Corresponding author}
\ead{ai.zhuk2@gmail.com}

\address[a]{Department of Mathematics and Physics, North Carolina Central University, \\1801 Fayetteville St., Durham, North Carolina 27707, U.S.A.}
\address[b]{Center for Advanced Systems Understanding,\\ Untermarkt 20, 02826 G\"{o}rlitz, Germany}
\address[c]{Helmholtz-Zentrum Dresden-Rossendorf,\\ Bautzner Landstra\ss e 400, 01328 Dresden, Germany}
\address[d]{Department of Physics, Istanbul Technical University,\\ 34469 Maslak, Istanbul, T\"{u}rkiye}
\address[e]{Astronomical Observatory, Odessa I.I. Mechnikov National University, \\ Dvoryanskaya St. 2, Odessa 65082, Ukraine}

\begin{abstract}
We study the effect of box size on the matter power spectrum obtained via cosmological N-body simulations. Within the framework of the cosmic screening approach, we show that the relative deviation between the spectra for our largest comoving box with $L=5632$ Mpc/$h$ and those for $L= 280, 560, 1680, 4480,5120$ Mpc/$h$ boxes consistently increases with decreasing box size in the latter set in the redshift range $0\leq z\leq 80$ for the considered values. As an additional demonstrative example, at redshift zero, we determine the values $k_{1\%}$ corresponding to the modes at which relative deviations reach 1\%.

\end{abstract}


\end{frontmatter}

\

\section{Introduction}
\label{sec:introduction}

The large-scale structure of the Universe is studied via both analytical methods \cite{Mukhanov2, Rubakov} and numerical simulations. It is well known that the former are appropriate in the linear regime, i.e. for small fluctuations in the matter density, and that such conditions apply to early epochs in the cosmic evolution, as well as large-enough domains in the late Universe. Using numerical simulations, regions with both small and large density contrasts can be explored, covering distances from astrophysical to cosmological scales, and of particular interest are relativistic simulations based on the complete system of Einstein's equations. These allow to straightforwardly include the relativistic effects that arise at large cosmological scales (e.g., the presence of a horizon, modifications to the gravitational interaction), consider objects with relativistic velocities such as neutrinos and study the influence of backreaction on the metric from perturbations. In this connection, the behaviour of scalar and vector potentials as well as of the matter power spectrum was investigated within the framework of the \textit{screening} formalism \cite{Eingorn1,fluids2,fluids3} in the recent papers \cite{vs1,scrmassdens}, using the relativistic N-body code \textit{gevolution} \cite{gevNat,gevolution}. First, a comparative analysis was provided in \cite{vs1}, which showed  that the spectra of metric perturbations are in good agreement between \textit{screening} and the default formulation of \textit{gevolution}, regardless of the differences in their weak-field expansion schemes. In \cite{scrmassdens}, the matter power spectrum obtained for the screening approach, in a comoving box of 5.632 Gpc/$h$, has confirmed that the temporal growth of density contrasts is suppressed at scales exceeding the Yukawa-type cutoff for the gravitational potential, as a manifestation of the so-called \textit{cosmic screening effect}.

In the present paper,  we investigate the effect of box size on the matter power spectrum within the cosmic screening approach. Due to the finite size of the box and periodic boundary conditions, modes exceeding the box size $L$  are not taken into account and the power spectrum is truncated at the fundamental mode $k_{\mathrm{fund}}=2\pi/L$  \cite{BS2,BS5,BS10,BS11} in N-body simulations. To investigate the associated effects quantitatively, we have performed six runs in boxes with comoving sizes ranging from 280 to 5632 Mpc/$h$, and with 2 Mpc/$h$ resolution. Here we study the resulting spectra for the redshifts $z= 80, 50, 15,0$, and as a suitable parameter that would characterize the effect of box size, obtain the modes $k_{1\%}$ for $z=0$, beyond which the relative deviation between the power spectrum of the largest box and the box under consideration becomes greater than 1$\%$.

The remainder of the paper is organized as follows: in Section 2, we present the power spectrum of the mass density contrast within the cosmic screening approach, obtained in N-body simulations of six different box sizes. We study the relative deviations for the spectra obtained for the 5632 Mpc/$h$ box vs. five smaller boxes, ranging from 280 Gpc/$h$ to 5120 Gpc/$h$ in comoving size, for $z\leq 80$. We briefly summarize the results in Section 3.

\

\section{N-body simulations of the matter power spectrum}
\label{sec:gevolution}

The power spectrum $P_{\delta}(k,z)$ of matter density contrast $\delta({\bf r},z)\equiv \delta\rho({\bf r},z)/\bar\rho = \left[\rho({\bf r},z)-\bar\rho\right]/\bar\rho$ may be defined by the equation \cite{gevNat,Durrer}
\be{1}
4\pi k^3\langle \widehat\delta({\bf{k}},z)\, \widehat\delta({\bf{k}}',z)\rangle =(2\pi)^3\delta_{\rm D} ({\bf k}-{\bf k}')P_{\delta}(k,z)\, ,
\ee
where $\widehat\delta({\bf{k}},z)$ is the Fourier transform of $\delta({\bf{r}},z)$ and $\delta_{\rm D} ({\bf k}-{\bf k}')$ is the Dirac delta function.
Given a system of nonrelativistic masses $m_n$ with the respective comoving radius-vectors ${{\bf r}_n(\eta)}$ (and momenta ${\bf q}_n$), the mass density reads $\rho =\sum_n m_n \delta_{\rm{D}}({\bf r} -{\bf r}_n)$ and $\bar \rho$ corresponds to its average value. $P_{\delta}(k,z)$ is a dimensionless quantity. We note that the following definition is also often used in the literature (see, e.g., \cite{Rubakov}) for the power spectrum: $\tilde P_{\delta}(k,z)=(2\pi^2/k^3)P_{\delta}(k,z)$.   

Within the screening approach, the mass density fluctuation $\delta\rho$ follows from (see Eqs. (10) and (15) in \cite{vs1})
\be{2}
\Delta\Phi - 3\mathcal{H}\Phi' - 3\Phi\left(\mathcal{H}^2+4\pi G\bar\rho/a\right) =(4\pi G/a)\delta\rho\, ,
\ee
where $\Delta$ is the Laplace operator, $\Phi$ is the first-order scalar perturbation of the Friedmann-Lema\^{\i}tre-Robertson-Walker metric
in the conformal Newtonian gauge, $a$ is the scale factor, the prime denotes the derivative with respect to conformal time $\eta$, $\mathcal{H}\equiv(da/d\eta)/a$ is the conformal Hubble parameter  and $G$ is the gravitational constant.

To obtain the power spectrum of $\delta\rho$, we have used the relativistic code \textit{gevolution} \cite{gevNat,gevolution}, with modifications that implement the expansion scheme of the screening approach. That is, aside from the evolution equation for $\Phi$ given above, we have introduced $d({\bf q}_n)/d\eta=-m_n a\nabla \Phi$ as the equation of motion for nonrelativistic matter.    It is also important to point out that in the presence of the gravitational field, the power spectra of energy and mass density contrasts are no longer coincident and in the screening formalism, matter fluctuations are formulated in terms of the latter \cite{massen}, as shown explicitly in Eq.~\rf{2}. The curves obtained in our runs for the present paper are, therefore, those for the mass density contrast.

\begin{figure*}
	\resizebox{1.\textwidth}{!}{\includegraphics{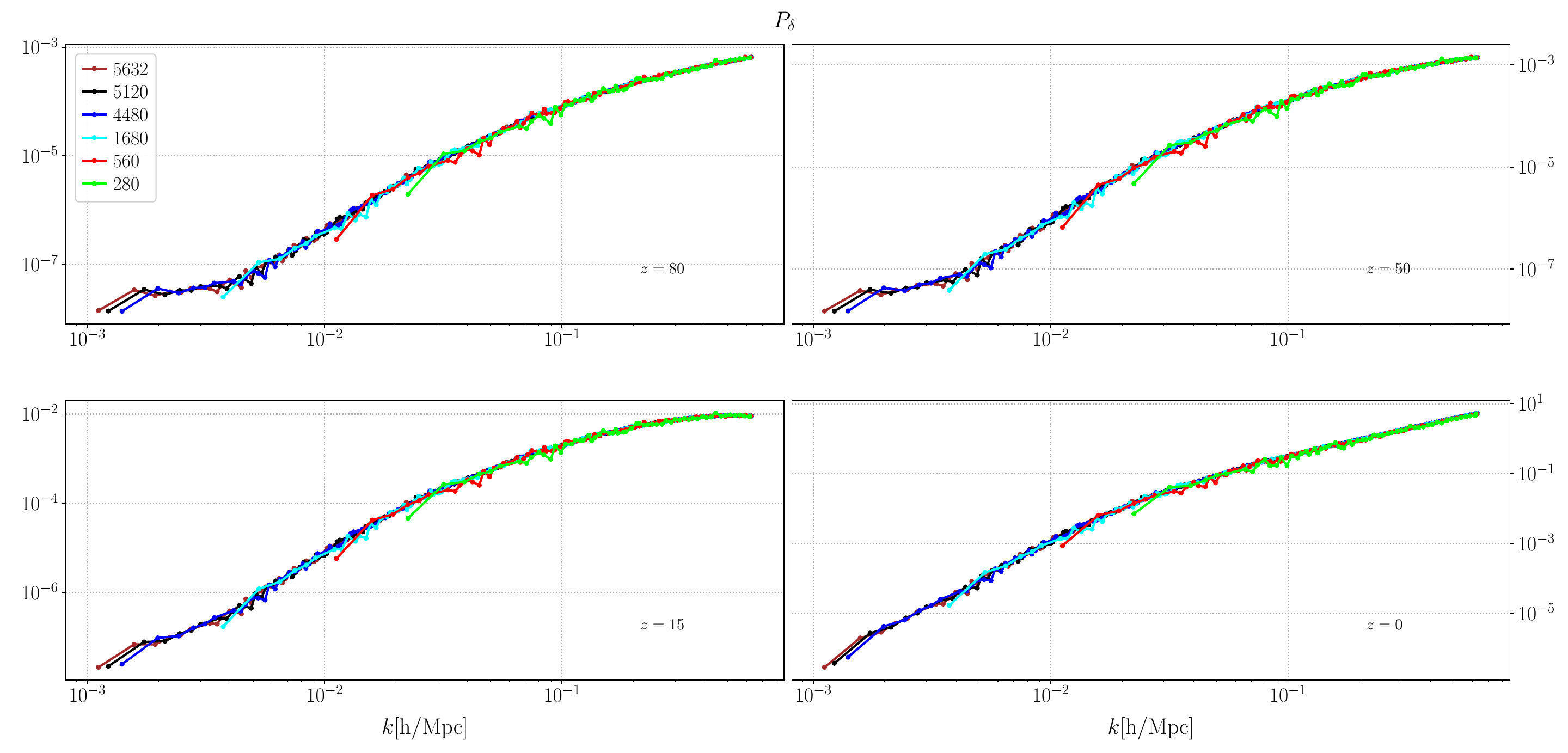}}
	\caption{Power spectrum $P_{\delta}(k,z)$ of the mass density contrast from Eq.~\rf{1} in boxes of 280, 560, 1680, 4480, 5120, 5632 Mpc/$h$ comoving size at  redshifts $z= 80, 50, 15,0$. The  parameters used to generate the curves were $H_0 = 67.66 \,  {\rm km}\, {\rm s}^{-1} {\rm Mpc}^{-1}$, $\Omega_{\rm M}h^2 = \Omega_{\rm b}h^2+\Omega_{\rm c}h^2=0.14175$, where $h=0.6766$,
and $\Omega_{\Lambda}=1-\Omega_{\rm M}$ \cite{planck2018} for $\Lambda$CDM cosmology.}
	\label{fig:1}
\end{figure*}

We have simulated $P_{\delta}(k,z)$ in boxes of $L= 280, 560, 1680, 4480, 5120, 5632$ Mpc/$h$ comoving size, with 2 Mpc/$h$ resolution. The results are presented in Fig.~\ref{fig:1} for $z= 80, 50, 15,0$. 
The ensemble of N-body particles consisted of nonrelativistic matter only, in which cold dark matter and baryons were treated on equal footing.   

\begin{figure*}
	\resizebox{1.\textwidth}{!}{\includegraphics{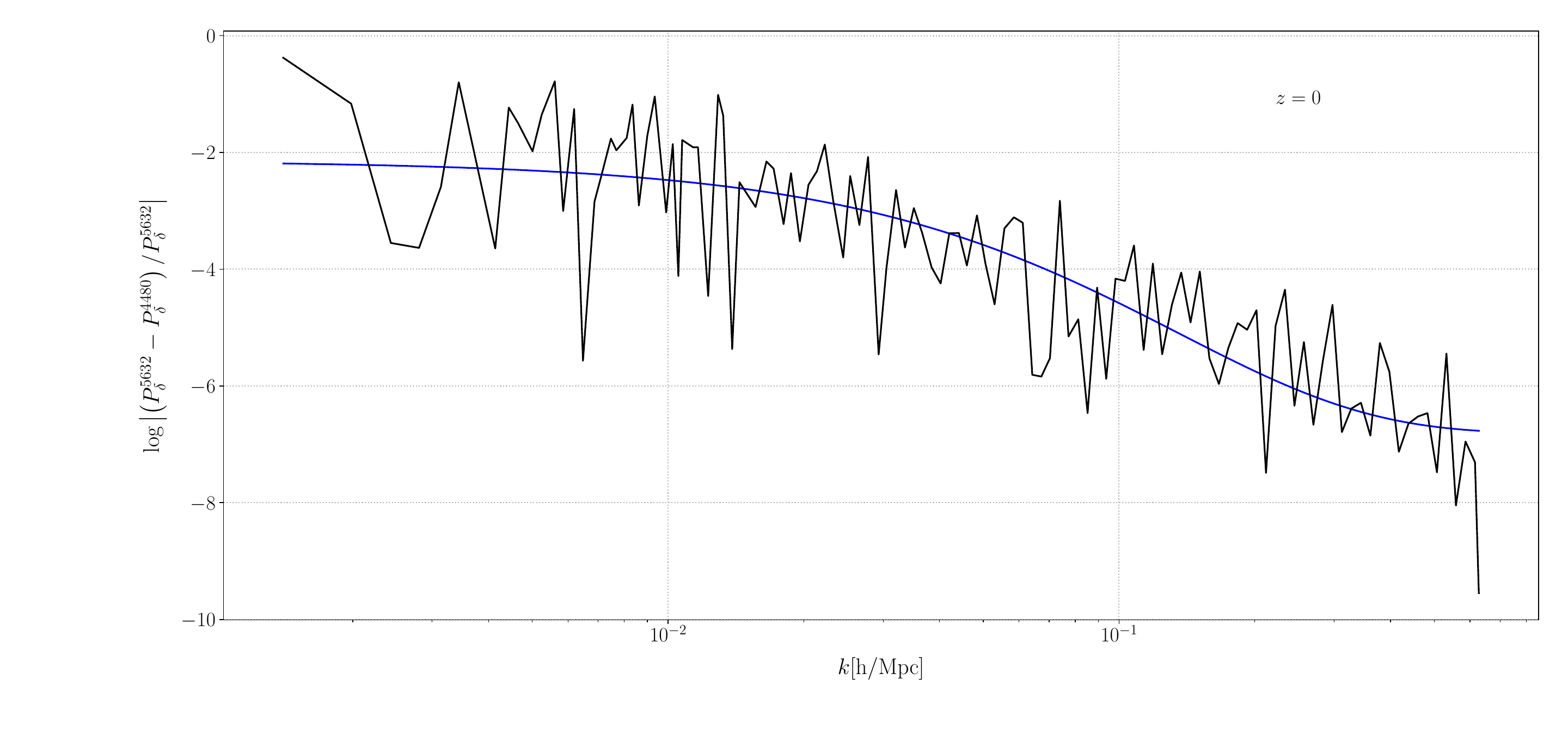}}
	\caption{Logarithm of the relative deviation defined in Eq.~\rf{3} at redshift zero for $X=4480$ (in black) and the corresponding smoothed curve (in blue).}
	\label{fig:2}
\end{figure*}

As expected, each curve in Fig.~\ref{fig:1} lies within a different range on the $x-$axis. The right edge (large k) is determined by the Nyquist limit, and it is the same for all curves, because for all boxes we have the same resolution. However, the left edge is defined by the fundamental mode $k_{\mathrm{fund}}=2\pi/L$,  which is size-dependent. It is clear that the larger $L$ is, the smaller $k_{\mathrm{fund}}$ becomes, and the finite box size has less impact on simulation results. To see this quantitatively, we introduce the relative deviation between the power spectrum for the largest box (in our case it is $L=5632$ Mpc/$h$) and those for smaller boxes as
\be{3}
\left|\frac{P^{5632}_{\delta}(k,z)-P^X_{\delta}(k,z)}{P^{5632}_{\delta}(k,z)}\right|\, ,
\ee
where $X$ is any of 280, 560, 1680, 4480, and 5120. In Fig.~\ref{fig:2}, we demonstrate the logarithm of this quantity for $X=4480$ at redshift zero. To highlight the general trend in the black curve, that has been generated directly from simulation outputs, and for demonstration purposes only, we smooth it (see the blue curve) using the function $a \exp(-b  x) + c$, for which the optimum parameters are estimated by minimizing the sum of squared residuals. We repeat the procedure for the remaining curves also, with the fitting parameters presented in Table~\ref{tab:abc}. 

\begin{table}[h]
	\centering
	\begin{tabular}{ c || c | c  }
	$L$ [Mpc/$h$]	&  $a,b,c$ & $a,b,c$  \\
		\hline\hline\\
		& $z=0$ & $z=15$\\
		\hline\\
		5120& 4.170, 11.502, -6.261 & 3.908, 18.614,  -5.799  \\
		4480&4.673,  7.385, -6.813 &  4.144, 16.638, -5.899   \\
		1680&  3.546, 10.813, -5.236 &  3.785,  8.004, -5.766  \\
		560& 2.697, 15.976, -3.630 &  3.529,  5.439, -4.993 \\
		280&  1.644, 10.449, -2.801 &  4.919,  0.973, -6.998  \\
		\hline\\
			&$z=50$ & $z=80$\\
	\hline\\
5120  &  3.811, 19.163, -5.737 &  3.820, 18.280, -5.775 \\
4480 &  4.095, 15.707, -5.914 & 4.079, 15.376, -5.929   \\
1680 & 3.799, 10.360, -5.588 &  3.843, 11.443, -5.545 \\
560 & 3.964,  3.983, -5.548 & 4.059,  3.819, -5.656 \\
280 &  2.432e3,  0.002, -2.434e3 &  4.379e3,  0.001, -4.381e3
	\end{tabular}
	
\caption{\label{tab:abc}
Fitting parameters for the model function $\; a \exp(-b  x) + c\; $ used to plot the curves in Fig.~\ref{fig:3}.}
\end{table}

\begin{figure*}
	\resizebox{1.\textwidth}{!}{\includegraphics{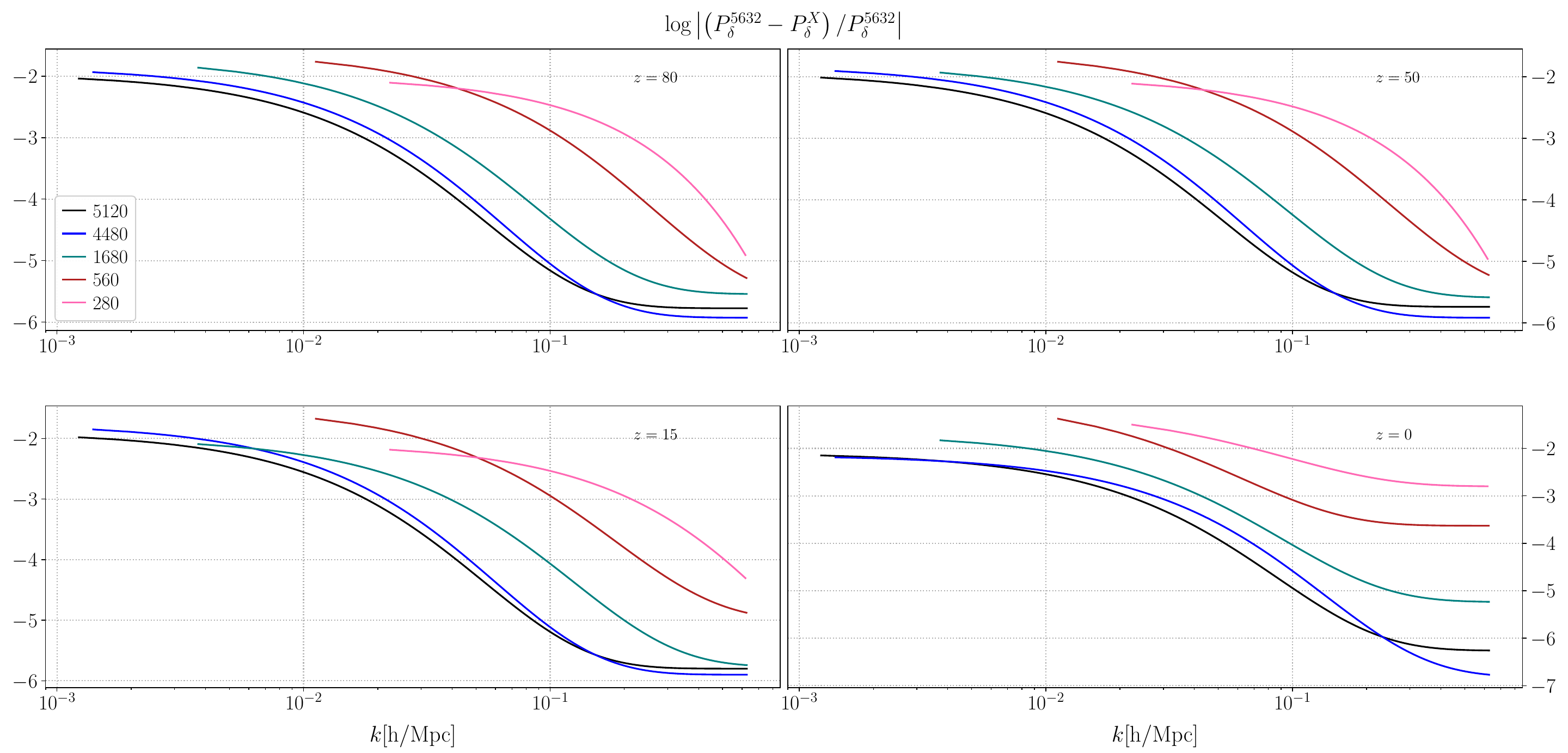}}
	\caption{Smoothed relative deviation curves for $X=280, 560,1680,$ $4480, 5120$ at redshifts $z=80, 50, 15$ and $z=0$.}
	\label{fig:3}
\end{figure*}

\begin{table}
\centering
	\begin{tabular}{c|c}
		$L$ [Mpc/$h$] & $k_{1\%}$\\
		\hline\\
		1680 & 0.00845 \\
		560 & 0.03155 \\
		280 & 0.06881
	\end{tabular}

\caption{\label{percent} The values of $k_{1\%}$ at which the relative deviation curves  in the lower right panel of Fig.~\ref{fig:3} (for $z=0$) reach 1\%. }
\end{table}

Relative deviations tend to increase with decreasing box size as confirmed by the behaviour of the smoothed curves in Fig.~\ref{fig:3}. To better demonstrate this growth, one can choose, for example, the  mode $k_{1\%}$ as a quantitative parameter, at which the  deviations reach 1\%. These are presented in Table~\ref{percent} for $z=0$. Since the curves are monotonically decreasing, the deviation is bigger than $1\%$ at all scales larger than $l_{1\%}=2\pi/k_{1\%}$. It is also worth highlighting that for boxes with sizes 4480 and 5120 Mpc/$h$, relative deviation remains less than 1\% for the entire range of the respective curves.

\section{Conclusion}
\label{sec:conclusion}

In the present paper we have considered the effect of the box size on the matter power spectrum obtained via cosmological  N-body simulations. We have conducted a series of runs in boxes of sizes  280, 560, 1680, 4480, 5120, 5632 Mpc/$h$ and obtained the  power spectrum $P_{\delta}(k,z)$ of the mass density contrast for each of them at four redshift values in the range $0\leq z \leq 80$. We have determined the relative deviations between the spectrum for the largest box vs. those for the five smaller boxes, which have shown an increasing trend with decreasing box size for all the redshift values investigated here. Quantitatively, this has been shown by defining the parameter $k_{1\%}$ for $z=0$, which determines the modes at which the relative deviations reach 1\%. Table~\ref{percent} demonstrates the increase in the value of $k_{1\%}$ with decreasing box size. 

\

\section*{Acknowledgments}

{\noindent This work was partially supported by the Center for Advanced Systems Understanding (CASUS) which is financed by Germany's Federal Ministry of Education and Research (BMBF) and by the Saxon state government out of the State budget approved by the Saxon State Parliament. Computing resources used in this work were provided by the National Center for High Performance Computing of T\"{u}rkiye (UHeM) under grant number 4007162019.}

\section*{Data Availability}

{\noindent The datasets generated in this study will be made publicly available in the Rossendorf Data Repository (RODARE).}

\

\end{document}